\documentclass[conference]{IEEEtran}
\IEEEoverridecommandlockouts
\usepackage{cite}
\usepackage{amsmath,amssymb,amsfonts}
\usepackage{algorithmic}
\usepackage{graphicx}
\usepackage{textcomp}
\usepackage{xcolor}
\usepackage{multirow}
\usepackage{hyperref}

\def\BibTeX{{\rm B\kern-.05em{\sc i\kern-.025em b}\kern-.08em
    T\kern-.1667em\lower.7ex\hbox{E}\kern-.125emX}}
\begin{document}

\title{Influence of Cell Position on the Capacity of Retired Batteries: Experimental and Statistical Studies \\
\thanks{This document is the results of the research project funded by the Région Auvergne-Rhône-Alpes.}
}

\author{
\IEEEauthorblockN{
 Marwan Hassini  \textsuperscript{1,2,3},Colette  Mintsa-Eya\textsuperscript{1},
Eduardo Redondo-Iglesias \textsuperscript{1,3}, 
 and Pascal Venet\textsuperscript{2,3}}
\IEEEauthorblockA{\textsuperscript{1}Univ Eiffel, ENTPE, LICIT-ECO7, 69500 Bron, France}
\IEEEauthorblockA{\textsuperscript{2} Université Claude Bernard Lyon 1, INSA Lyon, Ecole Centrale de Lyon, CNRS, Ampère, UMR5005,\\ 69100 Villeurbanne, France}
\IEEEauthorblockA{\textsuperscript{3}GEST (Eco7/Ampère Joint Research Team for Energy Management and Storage for Transport), 69500 Bron, France}
}
\maketitle

\begin{abstract}
Understanding how batteries perform after automotive use is crucial to determining their potential for reuse. This article presents experimental results aimed at advancing knowledge of retired battery performance. Three modules extracted from electric vehicles were tested. Their performance was assessed, and the results were analyzed statistically using analysis of variance (ANOVA). The 36 retired cells exhibited a high level of performance, albeit with significant variation. On average, the cells had a 95\% state of health capacity with a dispersion of 2.4\%. ANOVA analysis suggests that cell performance is not correlated with their position inside the module. These results demonstrate the need to evaluate dispersion within retired batteries and to develop thermal management and balancing systems for second-life batteries.\end{abstract}

\begin{IEEEkeywords}
Lithium-ion battery, Second life battery, Performance study
\end{IEEEkeywords}

\section{Introduction}

Reusing retired batteries is an interesting solution for reducing their environmental impact and limiting the extraction of raw materials needed to produce new batteries. After their automotive life, the performance of retired batteries must be assessed to determine their possible future uses \cite{hassini24}. This assessment is crucial for ensuring the viability of the second-life battery market. Developing rapid, cost-effective characterization techniques would streamline the repurposing process and improve the economic viability of reused batteries \cite{rallo2020}. Current state-of-the-art techniques rely on expensive equipment and require several hours of skilled labor\cite{hassini23}. These costs, coupled with the scarcity of batteries retired from automotive usage, result in limited knowledge of retired battery performance. This gap has two important consequences: the use of a theoretical threshold performance for retired batteries and an insufficient consideration of cell dispersion in battery performance assessments.

According to scientific literature, a capacity loss of 20-30\% is widely accepted as the end-of-life threshold for batteries \cite{haram21}. This theoretical threshold is rarely questioned, despite not being based on experimental data. Using this threshold has significant implications for research. In aging studies, for instance, this threshold is commonly used to determine when to stop testing, regardless of the battery technology or application \cite{jague15,pelos23}. Consequently, there is a wealth of experimental data on the aging of slightly degraded batteries. The models used to predict aging are particularly effective within this degradation range. However, knowledge is lacking for more severely degraded batteries. This results in less accurate estimates of degradation at low states of health. Studies on the cost and environmental impact of batteries extensively use the assumption of a fixed end-of-life threshold. This assumption negatively affects studies of environmental impact calculations and battery cost projections\cite{mathe20,korom22}.\\

Another important gap is the lack of experimental evaluation of dispersion inside retired batteries. Dismantling retired battery modules and characterizing each cell is a lengthy and complex process. For this reason, most studies of retired batteries have been conducted at the cell or assembly level without evaluating the performance of each cell. This leads to a knowledge gap regarding cell performance dispersion within the battery. Performance dispersion is commonly defined as the ratio of the standard deviation to the mean capacity value of the battery measurements \cite{schuster2015}. Understanding cell-to-cell performance variation is important for identifying potential weak spots in the module and determining which thermal and balancing systems to install. A cell at a weak spot would perform worse than its surroundings. As a first step, characterization of the retired module could be limited to assessing this cell, as it is well known that its performance will limit that of the entire module. A better understanding of the dispersion of retired batteries may also be useful for developing specific thermal and balancing systems for second-life batteries \cite{jeanneret2025,di2023novel}. Several authors have identified this issue as requiring particular attention in view of its impact on the performance of assemblies and on aging \cite{baumh14,seger22a}.\\

This study makes two significant contributions to existing literature. First, it presents capacity measurements for each cell in three retired modules. The results contribute to the limited number of existing studies on parameter dispersion in second-life batteries. The results improve the understanding of the primary issue of dispersion in battery packs. Additionally, this statistical study appears to be one of the earliest to apply analysis of variance to evaluate the impact of cell placement on the capacity degradation of retired batteries.

\section{Experimental study}
This section presents the characteristics of the batteries tested and the test conducted for this research work.

\subsection{Second life batteries tested}
For this experimental work, three battery modules were purchased from the second-life market. These modules came from BMW i3s with a 94-Ah capacity that were produced between 2016 and 2018. The tested BMW i3 module consists of 12 cells connected in series. The module was disassembled, and tests were conducted at the cell level. The SAMSUNG SDI 94 Ah cells are NMC111 with a prismatic format. Their main characteristics are summarised in table~\ref{table:samsung94Ahcharacteristics}.\\

\begin{table}[h!]
		\centering
	\caption{Main characteristics of the Samsung SDI 94~Ah cell.}
	\begin{tabular}{c c }
		\textbf{Characteristics} & \textbf{Values} \\
		\hline
		Format & Prismatic \\
		Nominal capacity [Ah] & 94 \\
		Nominal resistance [m$\Omega$] & 0.75 \\		
		Material at positive electrode & NMC111 \\		
		Material at negative electrode & Graphite \\	
		Nominal voltage [V] & 3.68 \\
		Voltage range [V] & 2.7-4.15\\
		Energy density [Wh/kg] &	165\\
		Dimensions L×W×H [mm] & 173×125×45 \\
		Mass [kg] & 2.1 \\
		\hline
	\end{tabular}
	\label{table:samsung94Ahcharacteristics}
\end{table}

\begin{figure}[h!]
	\centering
	\includegraphics[width=\linewidth]{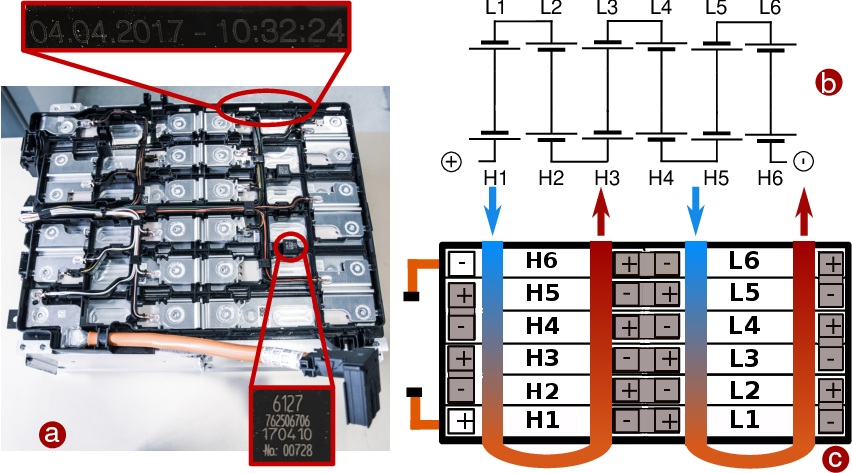}
	\caption{Distribution of cells in the module. Image (a) shows the module with the information of its production date. Image (b) shows the electrical connection between cells and images (c) shows how cells were named.}	\label{fig:cells_name_module}
\end{figure}

Figure~\ref{fig:cells_name_module} shows the organisation of a module. The cells are named according to their position in the assembly. The letter H or L distinguishes the cells according to their respective cooling circuit.

Table ~\ref{table:modules_tested} shows the production and purchase dates of the tested modules. The time elapsed between production and purchase enables estimation of the battery's automotive lifespan. Module 1 was used for less than four years in an electric vehicle, while modules 2 and 3 had a first life of around five years. According to the battery seller, modules 2 and 3 were extracted from the same electric vehicle.

\begin{table}[h!]
		\centering
	\caption{Dates of production, purchase and characterisation of the modules tested}.
	\begin{tabular}{cccc}		

		& \textbf{Manufacturer} & \textbf{Purchase}& \textbf{Characterisation }\\
		\hline
		\textbf{Module 1} & April 2017 & November 2020 & March 2022 \\
		\textbf{Module 2} & July 2016 & November 2021 & April 2023 \\
		\textbf{Module 3} & July 2016 & November 2021 & April 2023 \\
		\hline
	\end{tabular}
	\label{table:modules_tested}
\end{table}

\subsection{Test Protocol}

\begin{figure}[h!]
	\centering
	
	\includegraphics[width=\columnwidth]{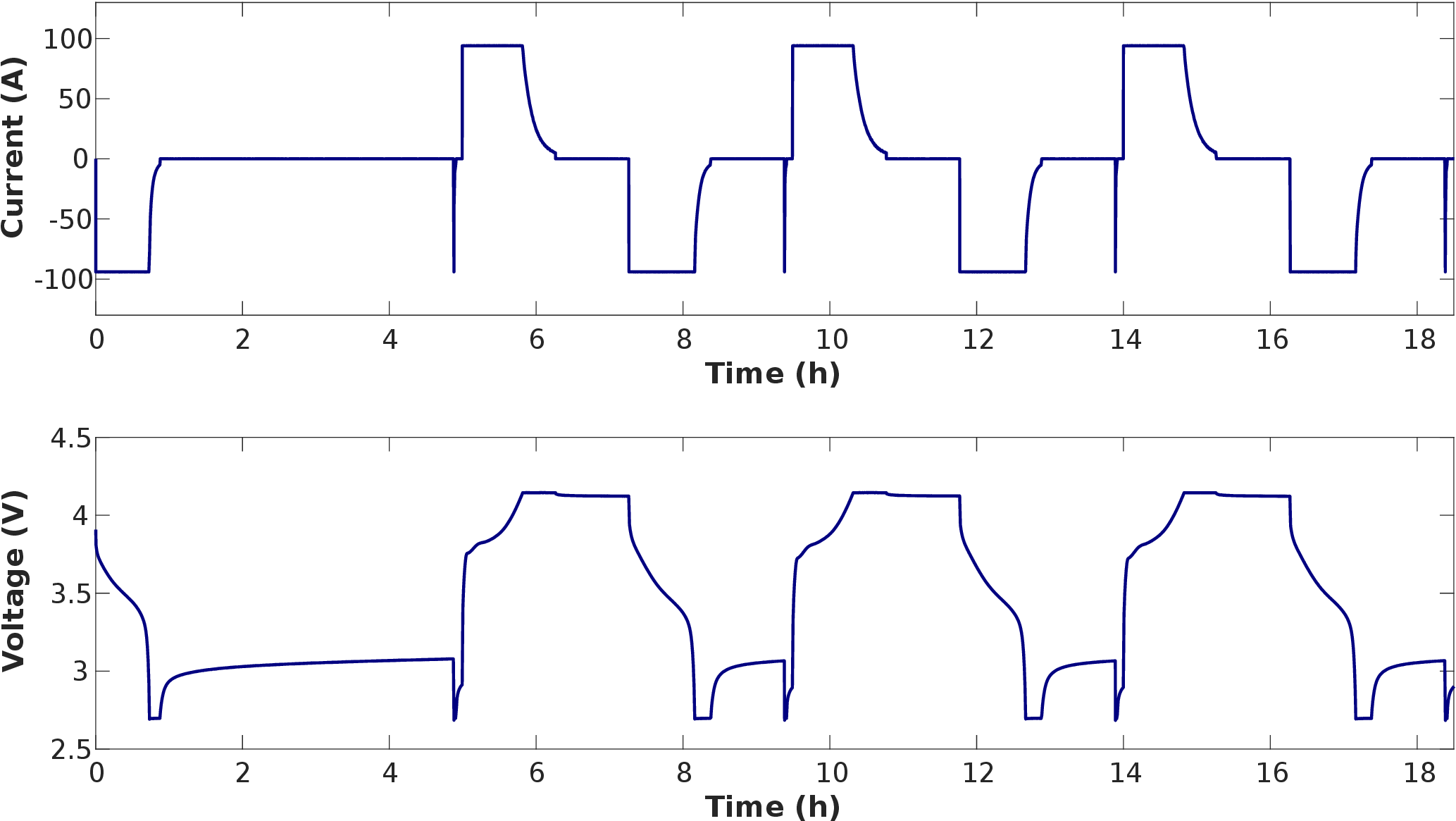}
	\caption{Evolution of current and voltage during the capacity test}.	
	\label{figure:profile_tension_carac_ref_capacity}
\end{figure}

The capacity measurement is made at 25°C. The test is illustrated in figure ~\ref{figure:profile_tension_carac_ref_capacity}.

The first step is to discharge the cell to minimum voltage. This is followed by a four-hour rest period. This initial rest period allows the temperature in the environmental chamber to reach the set temperature. Next, a sequence of three complete charge/discharge cycles is performed. Before each charge, a brief discharge restores the battery to its minimum voltage level. Then, charging occurs between the cell's minimum and maximum voltage thresholds by applying a constant current of 94 A (1 C). The maximum voltage is then maintained by reducing the current to 4.7 A (C/20). Discharge takes place between the same voltage thresholds at a current of -94 A (-1 C). The minimum voltage is maintained until the current reaches -4.7 A (C/20). This sequence is repeated three times. The first sequence serves as a preparation cycle, ensuring the cell temperature is close to that of the environmental chamber. This is useful when the cell has been stored at a different temperature than the characterization temperature. The next two cycles ensure repeatable measurements. For all tests conducted during this study, the variation between measurements in these two cycles is less than 0.5\%. The third capacity measured is discussed throughout the rest of the article.

\newpage
\section{Measurement results}
This section presents the experimental results. All analyses in this section were performed using the free, open-source software DATTES \cite {dattes2023}. Evaluating the performance of all the cells that make up these modules makes it possible to assess the dispersion between the elements. The capacity test presented in the previous section was performed on all cells within the module. Table ~\ref{tab:dispersion_capacite_3_modules_Qch} shows the capacity measurements for each cell.
\begin{table}[h!]
    \centering
        \caption{Cell capacity measurements (in amperhour) on three modules (M1, M2, M3). Data organised by cell position inside each module (H1-6, L1-6)}
    \begin{tabular}{c|ccc||c|ccc}
         & M1 & M2 & M3 & &M1 & M2 & M3 \\
         \hline
         H1 & 93.8&91.6 &91.0 &  L1 & 93.4 & 88.4 & 89.5\\
         H2 & 92.6 & 88.1 & 89.2 & L2 & 91.5 & 88.3 & 90.8\\
         H3 & 93.8 & 87.6& 88.9 & L3 & 92.8 & 90.1 & 90.1 \\
         H4 & 91.9 & 88.5 & 89.2 & L4 & 92.5 & 90.2 & 89.7\\
         H5 & 84.3 & 88.6 & 88.6 & L5 & 94.2 & 89.2 & 89.8\\
         H6 & 92.4 & 90.3 & 89.1& L6 & 93.4 & 88.6 & 90.2\\
         \hline
    \end{tabular}
    \label{tab:dispersion_capacite_3_modules_Qch}
\end{table}

\begin{figure}[h!]
	\centering
	\includegraphics[width=\columnwidth]{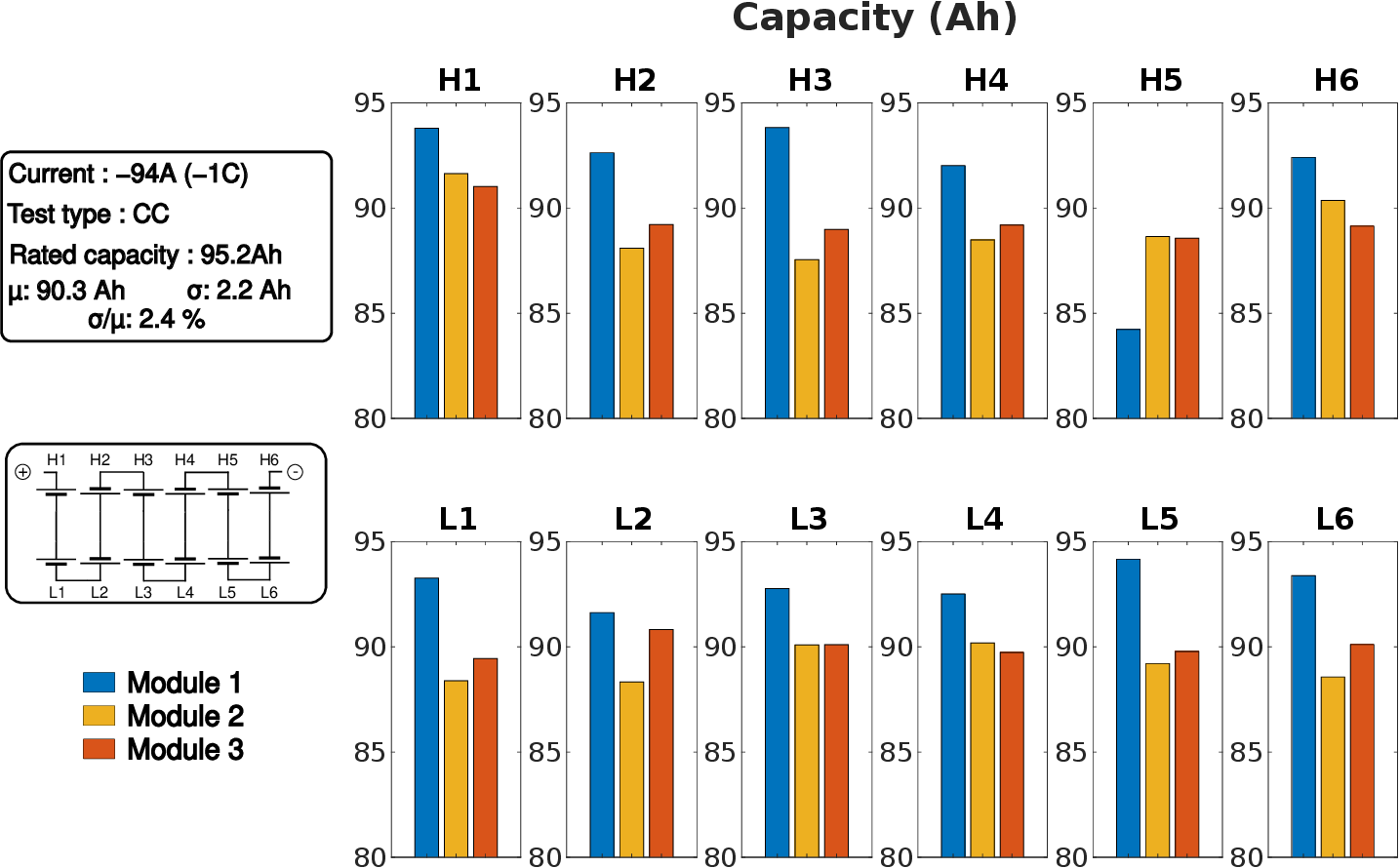}
	\caption{Capacity measurements on the three modules.}	\label{fig:soh_capacity_m1_m2_m3}
\end{figure}

In the datasheet, the rated capacity measured under comparable conditions is 95.2 Ah. The 36 retired cells maintained high performance, with an average state of health capacity of 95\%. These results demonstrate that, even after several years of automotive use, a battery can maintain excellent performance. This demonstrates that the end of the first life cycle is not solely related to performance degradation. Examples of reasons that can lead to the end of first life include accidents, driver decisions, and regulations. This level of performance is well above the commonly accepted second-life threshold of 70-80\%. These results suggest that defining automotive end-of-life based on experimental results and probability density would be more relevant. For the tested modules, the capacity dispersion is 2.4\%. This level of dispersion is consistent with the order of magnitude of capacity measurements in other studies involving batteries from electric vehicles \cite{jiang2017, braco2021experimental}. Figure~\ref{fig:soh_capacity_m1_m2_m3} shows the results of the capacity measurements as a function of the cells' position in the module.\\
\newpage

This figure illustrates the dispersion of capacity measurements across the three modules. The cells in Module 1 generally have a higher capacity, except for Cell H5, which significantly impacts the performance of this module due to the series connection. Modules 2 and 3 have similar performance levels. Their capacities and dispersions are similar. This corroborates the fact that these two modules are derived from the same electric vehicle. 

These measurements also show that there are no weak spots in the modules. In other words, no single cell position experiences higher degradation across all modules simultaneously. For instance, cell H5 in Module 1 is significantly degraded, whereas cells in the same position in Modules 2 and 3 perform similarly to the others. Section~\ref{section:statistical_study} will carry out a statistical study of the relationship between performance and position to confirm this observation.

\section{Influence of position on performance}~\label{section:statistical_study}

An interesting approach to quickly and economically determining the performance of a retired battery would be to identify the critical cell positions within a module. If a cell's position in a module significantly impacts its performance, then testing only the cells in these positions could reduce battery characterization time. However, the graphs in the previous section indicated the absence of such positions. This study aims to confirm or refute this observation through statistical analysis.

\subsection{ANOVA method}

To achieve this objective, the relationship between a cell's position in a module and its performance level is analyzed. This study uses the analysis of variance (ANOVA) method. ANOVA is used to explain the influence of a qualitative variable, called a factor, on a quantitative variable, the variable of interest. In this study, the factor is position, and the variable of interest is capacity. This method tests the following hypotheses:
\begin{itemize}
	\item[$\bullet$] Null hypothesis ($Hyp_0$): the factor has no influence on the variable of interest, i.e. no statistical difference between the means of the different levels.
	\item[$\bullet$] Alternative hypothesis ($Hyp_1$): the factor has an influence on the variable of interest, i.e. for at least one group, the mean is statistically different from the other means.
\end{itemize}

The ANOVA conclusion is based on the p-value, or probability value. This value describes the likelihood of obtaining a result that supports the null hypothesis. The risk threshold is conventionally set at 0.05 (5\%). If the p-value is less than 0.05, the null hypothesis is rejected. This indicates that the studied factor influences the variable of interest. Conversely, if p is greater than 0.05, the null hypothesis is not rejected, meaning the factor has no influence on the variable of interest.

\subsection{ANOVA with different subgroups}

\begin{figure*}[h!]
	\centering
	\includegraphics[width=\linewidth]{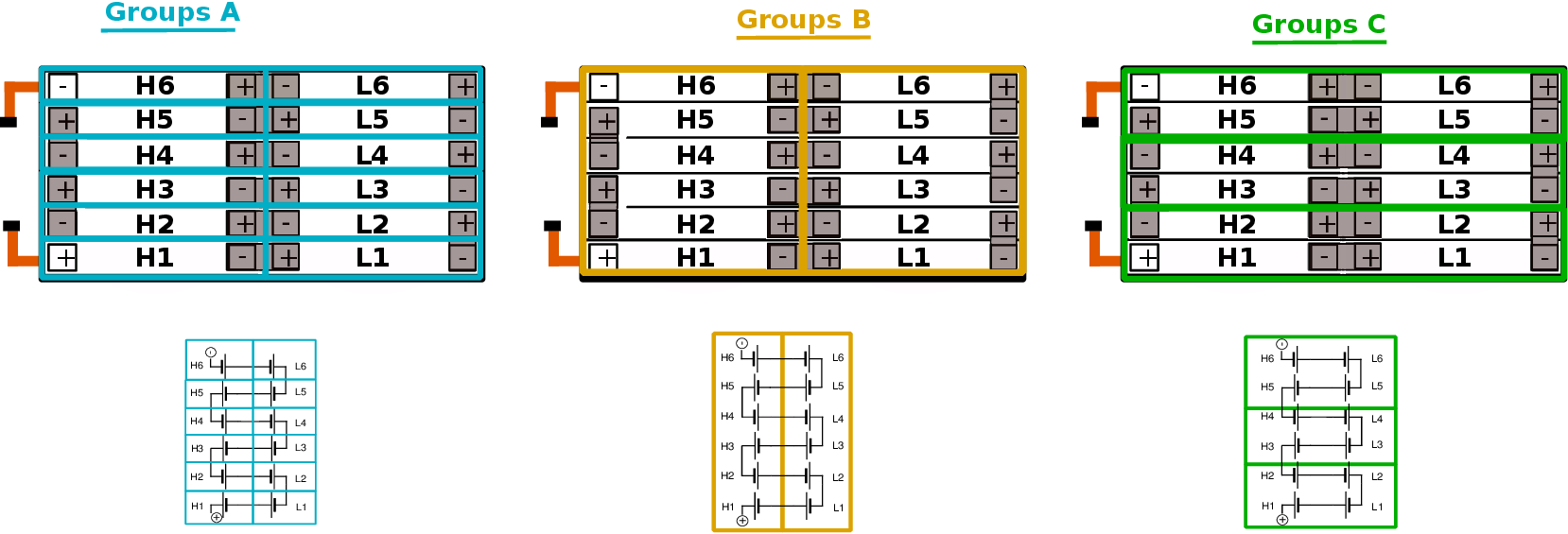}
	\caption{Grouping of cells studied in ANOVA. These groups show how each module was abstractly decomposed in groups. Each organization was repeated for each module used.}
	\label{figure:3_organisations_anova}
\end{figure*}

The ANOVA study was conducted on the groups of positions shown in figure~\ref{figure:3_organisations_anova}. 

\newpage

Configuration A (Group A) enables the study of positional influence by defining one group for each cell position. Using the three modules, 12 groups of three measurements each can be defined. Configuration B (Groups B) allows one to study the influence of position by defining two groups along a longitudinal section corresponding to cooling flows. In a vehicle, cells whose name begins with "H" are located on the vehicle's central axis. Cells beginning with L are located outside the battery pack. Using the three modules, two groups of 18 measurements each can be defined. Configuration C (groups C) allows one to study the influence of position by defining groups according to the proximity of cells within the module. To study the influence of cell position on cell capacity, the Matlab function 'anovan' was used \cite{matlab_anovan}. Table~\ref{tab:results_synthesis} shows the groups of cells made in each configurations and p-values obtained for the different analyses.\\

\begin{table}[h]
		\caption{Statistical analysis of capacity measurements in the three retired modules.}
        \label{tab:results_synthesis}
\begin{tabular}{cccc}
Group type  & Cells in group    & Mean capacity (Ah) & p value                \\\hline
\multirow{12}{*}{GROUPING A} & H1                & 92.1    & \multirow{12}{*}{0.62} \\
                             & H2                & 90.0    &                        \\
                             & H3                & 90.1    &                        \\
                             & H4                & 89.9    &                        \\
                             & H5                & 87.2    &                        \\
                             & H6                & 90.6    &                        \\
                             & L1                & 90.4    &                        \\
                             & L2                & 90.3    &                        \\
                             & L3                & 90.7    &                        \\
                             & L4                & 90.8    &                        \\
                             & L5                & 91.0    &                        \\
                             & L6                & 90.7    &                        \\\hline
\multirow{2}{*}{GROUPING B}  & H1-H2-H3-H4-H5-H6 & 90    & \multirow{2}{*}{0.37}  \\
                             & L1-L2-L3-L4-L5-L6 & 90.6    &                        \\\hline
\multirow{3}{*}{GROUPING C}  & H1-H2-L1-L2       & 90.7    & \multirow{3}{*}{0.67}  \\
                             & H3-H4-L3-L4       & 90.4    &                        \\
                             & H5-H6-L5-L6       & 89.9    &      \\             
\end{tabular}
\end{table}

The p-value is greater than 0.05 in all group configurations. The null hypothesis cannot be rejected based on the measured data. Therefore, it appears that the position of the cells does not affect the measurement of their capacity. This result is consistent with observations made in other articles on second-life batteries \cite{schuster2015,braco2021experimental}. \\

The results suggest that the thermal management and balancing systems maintained uniform temperatures and states of the cells in the modules. This mitigates the detrimental effects of high temperatures and dispersion on degradation. These results underscore the importance of cooling and balancing systems in enabling the reuse of batteries.

However, three retired modules is an insufficient sample size from which to draw further conclusions. The results of this study only allow us to identify a trend. The study suggests that the position of cells within an assembly is not a reliable indicator of performance. Evaluating retired batteries by assessing cells at weak spots is insufficient. Instead, characterization techniques that can quickly estimate battery performance and dispersion are necessary. Further experimental and statistical studies on other types of retired batteries with different states of health could complement this work well.

\section{Conclusions and outlook}
This article presents experimental results aimed at improving knowledge of retired battery performance. The cells of three modules extracted from electric vehicles were tested. The 36 retired cells exhibited high performance levels, albeit with variation. On average, the cells had a 95\% state-of-health capacity, with a dispersion of 2.4\%. This level of performance far exceeds the commonly accepted theoretical threshold of 70-80\% for a batteries second life. These results suggest that defining end of first life based on experimental results and probability density would be more relevant. This work also highlights the need to study the dispersion of retired batteries.\\

An analysis of variance also showed that position is not correlated with performance. This finding suggests that the thermal management and balancing systems maintained uniform temperatures and states of the cells in the modules. These results underscore the importance of cooling and balancing systems in enabling the reuse of batteries.\\

Future work could entail extending these results to other retired batteries and developing thermal and balancing strategies to limit the degradation and dispersion of reused batteries.





\vspace{12pt}

\end{document}